\documentclass[12pt]{article}
\usepackage{amsmath,amssymb,epsfig,color, url}
\usepackage{graphicx}
\usepackage{bm}

\def\slash#1{#1\!\!\!/\!\,\,}

\begin{document}

\begin{titlepage}

\begin{flushright}
UdeM-GPP-TH-15-245\\
WSU-HEP-1506\\
October 19, 2015
\end{flushright}

\vspace{0.7cm}
\begin{center}
\Large\bf 
Model-independent determination of the axial mass parameter
in quasielastic antineutrino-nucleon scattering\end{center}

\vspace{0.7cm}
\begin{center}
{\sc  Bhubanjyoti Bhattacharya$^{(a)}$,  Gil Paz$^{(b)}$, Anthony J. Tropiano$^{(c)}$}\\
\vspace{0.4cm}
{\it $^{(a)}$ Physique des Particules, Universit\'e de Montr\'eal, \\ C.P. 6128, succ.\ centre-ville, Montr\'eal, QC, Canada H3C 3J7
}\\
\vspace{0.3cm}
{\it 
$^{(b)}$ 
Department of Physics and Astronomy \\
Wayne State University, Detroit, Michigan 48201, USA 
}\\
\vspace{0.3cm}
{\it 
$^{(c)}$ 
Department of Physics and Astronomy\\ 
Michigan State University,
East Lansing, Michigan 48824, USA}

\end{center}
\vspace{0.7cm}

\begin{abstract}
  \vspace{0.2cm}
  \noindent
Understanding the charged current quasielestic (CCQE) neutrino-nucleus interaction is important for precision studies of neutrino oscillations. The theoretical description of the interaction depends on the combination of a nuclear model with the knowledge of form factors. While the former has received considerable attention, the latter, in particular the axial form factor, is implemented using the historical dipole model. Instead, we use a model-independent approach, presented in a previous study, to analyze the muon antineutrino CCQE mineral oil data published by the MiniBooNE collaboration. We combine the cross section for scattering of antineutrinos off protons in carbon and hydrogen, using the same axial form factor for both. The extracted value of the axial mass parameter $m_A = 0.84^{+0.12}_{-0.04} \pm {0.11}  \, {\rm GeV}$ is in very good agreement with the model-independent  value extracted from MiniBooNE's neutrino data. Going beyond a one-parameter description of the axial form factor, we extract values of the axial form factor in the range of $Q^2=0.1...1.0$ GeV$^2$, finding a very good agreement with the analogous extraction from the neutrino data. We discuss the implications of these results.

\end{abstract}
\vfil

\end{titlepage}

\section{Introduction}
Future neutrino oscillation experiments plan to study the neutrino-mass ordering and search for CP violation in the lepton sector of the standard model. In order to do that, the charged current quasielestic (CCQE) neutrino-nucleus interaction must be known to high precision, see e.g. \cite{AguilarArevalo:2013hm}.

The neutrino interaction with quarks is determined by the standard model Lagrangian.  For neutrino-nucleus scattering this interaction is folded twice. First, going from the quark level to the nucleon level form factors must be introduced. Second, going from the nucleon level to the nucleus level a nuclear model must be introduced. Thus neutrino-nucleus interaction is determined by the \emph{combination} of form factors and a nuclear model. To understand neutrino-nucleus interaction it is important to get a handle on both. 

The issue of whether the nuclear models used by the neutrino experiments are adequate was discussed by many authors. The question of the form factors has received a lot less attention \cite{Bhattacharya:2011ah}. In the isospin limit there are four form-factors that contribute to the interaction: $F_1, F_2, F_P$, and $F_A$, see the appendix for details. Two of these, $F_1$ and $F_2$, can be related to the electric and magnetic form factors extracted in electron-proton scattering.  $F_P$ can be related to $F_A$ in the $m_\pi\to 0$ limit, using PCAC. Furthermore, its contribution to the cross section is suppressed by $m^2_\ell/m_N^2$ where $m_\ell$ is the charged lepton mass and $m_N$ is the nucleon mass, see the appendix. Even if the charged lepton is a muon, $m^2_\ell/m_N^2$ is only about 1\%. Its contribution is further suppressed for an electron. 

That leaves the axial form factor $F_A(q^2)$. Its value at $q^2=0$ can be determined from neutron decay. In particular, the latest value from the particle data group is $F_A(0)=-1.272$  \cite{Agashe:2014kda}.   A common model used in the literature to parametrize $F_A$ is the dipole model, 
\begin{equation}\label{dipole}
F_A^{\rm dipole}(q^2) = {F_A(0) \over \left[  1 - {q^2/ (m_A^{\rm dipole})^2}  \right]^2 }. 
\end{equation}
There are several problems with this model. First, it is motivated by similar older dipole models of the electromagnetic form factors. These are now known to be inadequate to describe electromagnetic form factor data. There is no reason to believe the dipole model is  adequate to describe $F_A$. Second, this is a one-parameter model.  One should not expect \emph{any} one-parameter model to always adequately describe the form factor.  As neutrino data becomes more accurate, the only improvement possible in this model is to reduce the error on $m_A^{\rm dipole}$. Third, it is not clear what the physical meaning of $m_A^{\rm dipole}$ is. When different  extractions of  $m_A^{\rm dipole}$ disagree, is it a real discrepancy in the data or is it an artifact of the use of the dipole model?  One would like to have a general improvable parametrization with its parameters having a model-independent interpretation.    

The solution to this problem lies in using the so called $z$ expansion for $F_A$ \cite{Bhattacharya:2011ah}. This method relies on the known analytic properties of the form factor to express it as a Taylor series in the variable $z(q^2)$, i.e. $F_A(q^2)=\sum_k a_k z^k(q^2)$ (see section \ref {sec:theory} below). The $z$ expansion has by now become the standard tool in analyzing meson form factors, see e.g. \cite{Aoki:2013ldr}. It was first applied to baryons, in particular the proton electric form factor in \cite{Hill:2010yb}. It has also been applied successfully to extract the nucleon axial mass in \cite{Bhattacharya:2011ah}, the proton and the neutron magnetic radii in \cite{Epstein:2014zua}, 
the proton electric and magnetic radii in \cite{Lorenz:2014yda} (using also other methods), and the proton electric and magnetic radii in \cite{Lee:2015jqa}\footnote{Some other studies do not bound the coefficients of the $z$ expansion \cite{Lorenz:2014vha, Griffioen:2015hta} or modify it \cite{Horbatsch:2015qda}.}. It has also been applied to analyze heavy-baryon form factors in \cite{Khodjamirian:2011jp} and \cite{Detmold:2015aaa} and the strange nucleon electromagnetic form factors in \cite{Green:2015wqa}.

In particular \cite{Bhattacharya:2011ah} has used the $z$ expansion to extract axial mass in a  model-independent way. The axial mass is defined in terms of the form factor slope at $q^2=0$:  $m_A=\left[F_A^\prime(0)/ 2 F_A(0)\right]^{-1/2}$ \cite{Bhattacharya:2011ah}. Assuming the dipole model this definition coincides with $m_A^{\rm dipole}$. But in general the two are not equal. The extraction is model independent since the value of $m_A$ is independent of the number of parameters used in the fit. From the MiniBooNE muon-neutrino data \cite{AguilarArevalo:2010zc}, \cite{Bhattacharya:2011ah} found $m_A=0.85^{+0.22}_{-0.07} \pm {0.09}  \, {\rm GeV}$. This value is consistent with fits to an illustrative dataset for 
pion electroproduction,  $m_A = 0.92^{+0.12}_{-0.13}\pm 0.08 \,{\rm GeV}$. Using the dipole model,  \cite{Bhattacharya:2011ah} has found $m_A^{\rm dipole}=1.29\pm 0.05 \,{\rm GeV}$ (neutrino scattering) and 
$m_A^{\rm dipole}=1.00\pm 0.02 \,{\rm GeV}$ (electroproduction). One could conclude in this case that the discrepancy is an artifact of the use of the dipole model.

The results of  \cite{Bhattacharya:2011ah} assumed the Relativistic Fermi Gas (RFG) nuclear model
of Smith and Moniz~\cite{Smith:1972xh}. Still, it was possible to extract one of the RFG  parameters  $\epsilon_b$ from the MiniBooNE data without an assumption on $m_A$. In particular,  \cite{Bhattacharya:2011ah} has found $\epsilon_b = 0.028 \pm 0.03 \,{\rm GeV}$ in agreement with the value $\epsilon_b=0.025$ GeV, as extracted from electron scattering data on nuclei in \cite{Moniz:1971mt}, but less consistent with the value used by MiniBooNE \cite{AguilarArevalo:2010zc}, $\epsilon_b=0.034\pm0.09$ GeV. 

The agreement between the model-independent extraction of the axial mass from neutrino and pion electroproduction data is very encouraging.  It is important to confirm these results by applying the same method to other neutrino data sets. For example,  MiniBooNE has released data on muon antineutrino-nucleus scattering \cite{AguilarArevalo:2013hm}. An important difference from the neutrino-nucleus case is that the antineutrino scatters off protons in carbon and hydrogen, as opposed to only neutrons in carbon for the neutrino. As a result one has to combine scattering off ``free" protons in hydrogen with scattering off ``bound" protons in carbon. In particular,  \cite{AguilarArevalo:2013hm} has used \emph{different} values of the axial mass for protons in carbon ($m_A^{\rm dipole}=1.35$ GeV) and protons in hydrogen ($m_A^{\rm dipole}=1.02$ GeV). Since the axial mass is a fundamental property of the nucleon, such a treatment is problematic. The main goal of this paper is the model-independent extraction of the axial mass of the nucleon from the  antineutrino-nucleus scattering data, using the \emph{same} axial form factor for protons in carbon and hydrogen.

The paper is structured as follows. In section \ref {sec:theory} we briefly review the theoretical framework behind the model-independent extraction. In section \ref{sec:extraction} we present the results of model-independent extraction of the axial mass and the axial form factor from the data. We present our conclusions in section \ref{sec:summary}. For completeness, we have collected in the appendix formulas for the differential cross section for free and bound nucleon, including the effects of flux averaging.

\section{Theoretical Framework} \label{sec:theory}
Most of the theoretical framework concerning the use of the $z$ expansion was discussed in detail in \cite{Bhattacharya:2011ah}. Here we briefly review  it.  

The nucleon matrix elements depend on four form factors, $F_1, F_2, F_p,$ and $F_A$, see the appendix.  Two of them, $F_1$ and $F_2$, can be related using isospin symmetry to the electromagnetic form factors measured in electron-proton scattering. In extracting $m_A$ and $F_A$ from the MiniBooNE  data we generally try to follow their choices for the input functions and parameters. Thus we use the BBA2003 parameterization~\cite{Budd:2003wb} for $F_1$ and $F_2$, used in  \cite{AguilarArevalo:2013hm}. For $F_p$ we use the pion pole approximation\footnote{The values we use for the various input parameters are listed in Table \ref{tab:inputs}. In the following we assume that the errors due the variation of these parameters are small compared to the uncertainty on $F_A$. Also, due to the suppression of the $F_P$ contribution we will not consider uncertainties associated with this approximation.} $F_P(q^2) \approx 2 m_N^2 F_A(q^2)/  \left(m_\pi^2 -q^2\right)$. This leaves $F_A$ which is the focus of our analysis. 

The axial form factor is analytic in the cut $t=q^2$ plane outside a cut that starts at the three-pion threshold,  $t\ge t_{\rm cut}=9m_\pi^2$. The domain of analyticity can be mapped onto the unit circle via the transformation 
\begin{equation}
z(t,t_{\rm cut},t_0) = 
{\sqrt{t_{\rm cut}-t} - \sqrt{t_{\rm cut} - t_0}\over
\sqrt{t_{\rm cut}-t} + \sqrt{t_{\rm cut} - t_0}} \,, 
\end{equation}
where $t_0$ is a free parameter determined by $z(t_0,t_{\rm cut},t_0)=0$. In our analysis we will take $t_0=0$, but the results do not depend on this choice  \cite{Bhattacharya:2011ah}. We express the form factor as a power series in $z(q^2)=z(q^2,t_{\rm cut},0)$
\begin{equation}\label{Fz} 
F_A(q^2) = \sum_{k=0}^\infty a_k z(q^2)^k \,. 
\end{equation}
For $t_0=0$, $a_0$ is equal to $F_A(0)=-1.272$. The axial mass is determined from the slope of $F_A$ at $q^2=0$, i.e.  $m_A=\left[F_A^\prime(0)/ 2 F_A(0)\right]^{-1/2}$. For the choice of $t_0=0$, $m_A$ will depend only on $a_1$. To ensure that $m_A$ does not depend on the number of parameters, the coefficients must be bounded. As discussed in detail in \cite{Bhattacharya:2011ah}  we will use in our fits  the uniform bounds\footnote{It should be noted that the uniform bounds of 5, 10 also imply that $m_A>0.599,0.424$ GeV \cite{Epstein:2014zua}. In such cases one can perform fits with bounds on all $a_k$ apart from $a_1$. In practice, this is not a problem as the extracted values we find are larger than these lower bounds.} of  $|a_k|\le 5$ and $|a_k|\le 10$.    In practice we fit only a finite number of parameters $0<k\leq k_{\rm max}$. For definiteness our default is $k_{\rm max}=7$, but we have checked that fitting a larger number of parameters, i.e. $k_{\rm max}=8,9,$ and $10$ does not change the results.  

As in the MiniBooNE analysis we use the RFG nuclear model, but  similar to the analysis of \cite{Bhattacharya:2011ah}, we use the binding energy of $\epsilon_b=0.025$ GeV from \cite{Moniz:1971mt}.

The antineutrino can scatter off protons in carbon and in hydrogen. MiniBooNE reports the double differential cross section per nucleon for the mineral oil used in the experiment\footnote{MiniBooNE also reported ``hydrogen subtracted" data  \cite{AguilarArevalo:2013hm}. We do not use it for two reasons. First,  the subtraction relies on the event generator, since the scattering off carbon and hydrogen cannot be distinguished in the data. Second, it  uses a different axial mass for protons in hydrogen and protons in carbon.}. 
The mineral oil is composed of $C_nH_{2n+2}$, $n \sim 20$, see section IIIA of \cite{AguilarArevalo:2013hm}. This implies that on average there are 2.1 hydrogen nuclei for every carbon nucleus. Considering the large distance between the carbon and the hydrogen nuclei compared to the typical nuclear size, we can add the cross sections directly, ignoring interference, and divide by the number of protons, in this case 8.1. We thus have,
\begin{equation}\label{xsoil}
\frac{d\sigma_{\rm mineral\, oil,\, per\, nucleon,\, avg.}}{dE_\ell d\cos\theta_\ell}=\frac1{8.1}\left(6\frac{d\sigma_{\rm carbon,\, per\, nucleon,\, avg.}}{dE_\ell d\cos\theta_\ell}+2.1 \frac{d\sigma_{\rm hydrogen,\, avg.}}{dE_\ell d\cos\theta_\ell}\right). 
\end{equation}
where ``avg." denotes flux averaging. The expressions for  $d\sigma_{\rm hydrogen,\, avg.}$ and $d\sigma_{\rm carbon,\, per\, nucleon,\, avg.}$ are given in equations (\ref{xsfreeavg}) and  (\ref{xscarbonavg}).  In particular, we use the \emph{same} axial form factor for carbon and hydrogen cross sections.

The expression in (\ref{xsoil}) can be compared to MiniBooNE's reported per-nucleon mineral oil differential cross section to extract $m_A$. In order to do that we form the error matrix 
\begin{equation}
E_{ij} = (\delta \sigma_i)^2 \delta_{ij} + (\delta N)^2
\sigma_i \sigma_j  \,, 
\end{equation}
where $\sigma_i = (d\sigma/dE_\mu d\cos\theta_\mu)\Delta E_\mu \Delta \cos\theta_\mu$  denotes a partial cross section, $\delta \sigma_i$ denotes the shape uncertainty  from Table~XIV of \cite{AguilarArevalo:2013hm}, and $\delta N = 0.13$ is the normalization error from \cite{AguilarArevalo:2013hm}.  We form the chi-squared function  
\begin{equation}
\chi^2 = \sum_{ij} (\sigma^{\rm
expt.}_i-\sigma^{\rm theory}_i) E^{-1}_{ij}  (\sigma^{\rm
expt.}_j-\sigma^{\rm theory}_j) \,, 
\end{equation}
and minimize $\chi^2$ to find best fit values for $m_A$.  The error on $m_A$ is determined from the $\Delta \chi^2 = 1$ intervals. 

In order to study the $m_A$ sensitivity to $Q^2=-q^2$, we consider subsets of the MiniBooNE data with a cut on $Q^2$. For a free proton at rest, the $Q^2$ can be determined from the observed muon energy and scattering angle assuming a quasielastic scattering. Due to nuclear effects $Q^2$ cannot be determined unambiguously and as a proxy we use the ``reconstructed" $Q^2$ of \cite{Bhattacharya:2011ah},
\begin{equation}\label{Q2rec}
Q^2_{\rm rec}=2E^{\rm rec}_\nu E_\mu-2E^{\rm rec}_\nu\sqrt{E_\mu^2-m_\mu^2}\cos\theta_\mu -m_\mu^2 \,, 
\end{equation}
where $E_\nu^{\rm rec}$ approximates the neutrino energy in the nucleon rest frame, 
\begin{equation}
E^{\rm rec}_\nu= \frac{m_N E_\mu-m_\mu^2/2}{m_N-E_\mu+\sqrt{E_\mu^2-m_\mu^2}\cos\theta_\mu } \,. 
\end{equation}

\begin{table}
\begin{center}
\begin{tabular}{l|c|c}
Parameter & Value &  Reference \\
\hline
$|V_{ud}|$ & 0.9742 &  \cite{Agashe:2014kda} 
\\
$\mu_p$ & 2.793 & \cite{Agashe:2014kda} 
\\
$\mu_n$ & $-1.913$ & \cite{Agashe:2014kda} 
\\
$m_\mu$ & 0.1057 GeV & \cite{Agashe:2014kda} 
\\
$G_F$ & $1.166 \times 10^{-5}$ GeV$^{-2}$ & \cite{Agashe:2014kda} 
\\
$m_N$ & 0.9389 GeV & \cite{Agashe:2014kda} 
\\
$m_\pi$ & 0.140 GeV & \cite{Agashe:2014kda} 
\\
$F_A(0)$ & $-1.272$ & \cite{Agashe:2014kda} 
\\
$\epsilon_b$ & 0.025 GeV & \cite{Moniz:1971mt}
\\
$p_F$ & 0.220 GeV & \cite{AguilarArevalo:2010zc}

\end{tabular} 
\end{center}
\caption{\label{tab:inputs}
Numerical values for input parameters. 
} 
\end{table}

\section{Results}\label{sec:extraction}
\subsection{Axial mass extraction}

\begin{figure}
\begin{center}
\includegraphics[scale=1]{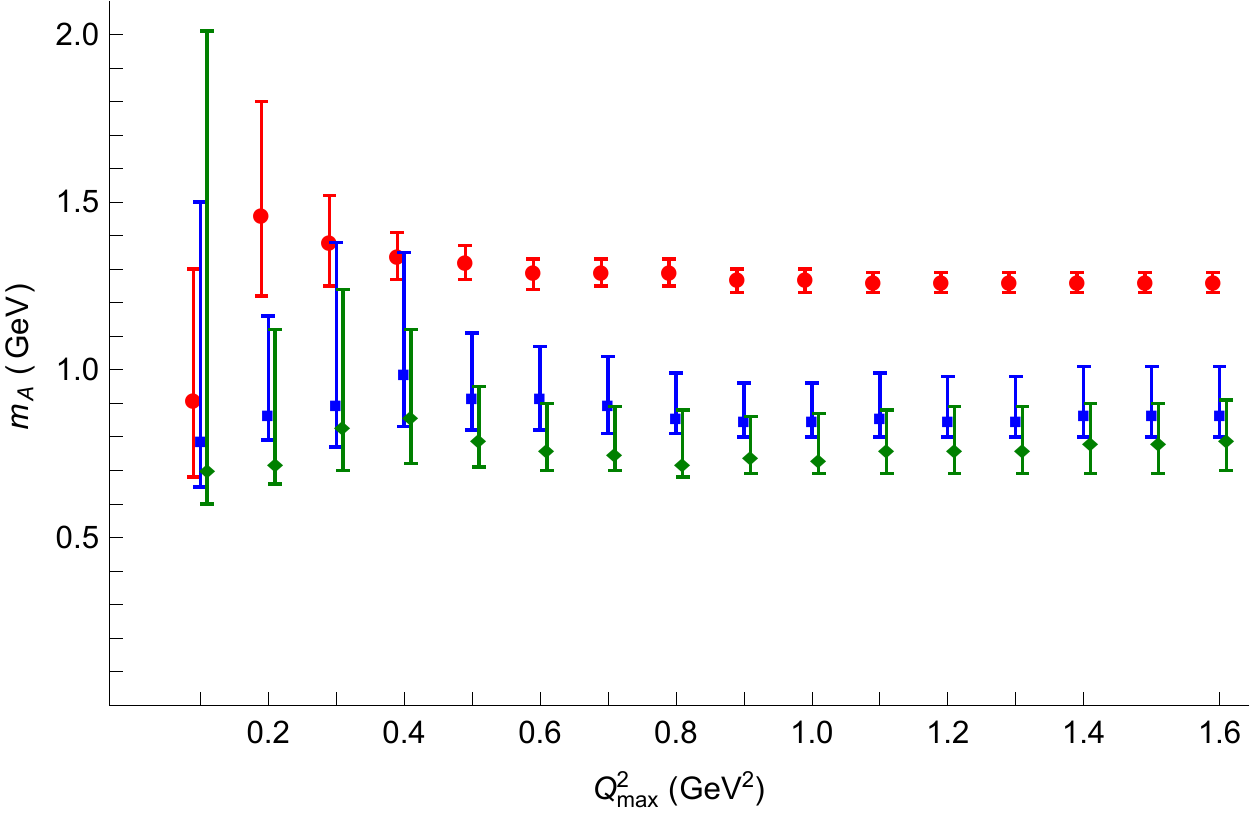}
\caption{\label{fig:mA_Q2} Extracted value of $m_A$ versus $Q^2_{\rm max}$. 
Dipole model results for $m_A^{\rm dipole}$ 
are shown by the red circles;  
$z$ expansion results with $|a_k|\le 5$ are shown by the blue squares, 
$z$ expansion results with $|a_k|\le 10$ are shown by the green diamonds.
}
\end{center}
\end{figure}  
As described in the previous section, $F_A(q^2)$ is the only undetermined part of  (\ref{xsoil}). We extract $m_A^{\rm dipole}$ using the dipole model from (\ref{dipole}) and $m_A$ using the $z$ expansion from (\ref{Fz}). We present results for data with $Q^2_{\rm rec}\leq Q^2_{\rm max}$, where $Q^2_{\rm max}=0.1, 0.2,..., 1.6$ GeV$^2$. The cut $Q^2_{\rm max}=1.6$ GeV$^2$ includes the entire MiniBooNE data set. We apply the bounds $|a_k|\leq 5$ and $|a_k|\leq 10$ and use $k_{\rm max}=7$. Our  results are presented in figure \ref{fig:mA_Q2}. The $z$ expansion results lie systematically below values extracted using the dipole model. The same effect was found in \cite{Bhattacharya:2011ah} analyzing MiniBooNE's  neutrino data. Taking $Q^2_{\rm max}=1.0$ GeV$^2$ for definiteness, we find 
\begin{equation}\label{eq:MAz}
m_A = 0.84^{+0.12}_{-0.04} \pm {0.11}  \, {\rm GeV}  \qquad \mbox{(antineutrino scattering)},
\end{equation}
where the first error is experimental, using the fit with $|a_k|\le 5$, and the second error represents residual form factor shape uncertainty, 
taken as the maximum change of the $1\sigma$ interval when the bound is increased to $|a_k|\le 10$. This result is in very good agreement with the value extracted in \cite{Bhattacharya:2011ah} from neutrino data 
\begin{equation}
m_A = 0.85^{+0.22}_{-0.07} \pm {0.09}  \, {\rm GeV}  \qquad \mbox{(neutrino scattering)}.
\end{equation}
For comparison, a fit that uses the dipole model and the same $Q^2_{\rm max}$ gives  $m_A^{\rm dipole} = 1.27^{+0.03}_{-0.04}$~GeV. This value is in good agreement with value found in \cite{Bhattacharya:2011ah} $m_A^{\rm dipole} = 1.29 \pm 0.05$~GeV. 

For both the neutrino and antineutrino data sets the errors on the model-independent axial mass are not symmetric. In particular the magnitude of the upper error bar is larger than the magnitude of the lower error bar. It was suggested \cite{McFarland} that it might be due to the fact that $m_A^2$ is inversely proportional to the slope $F_A$. We can check this hypothesis by extracting the ``axial radius" \cite{Bhattacharya:2011ah}, $r_A=\left[6F_A^\prime(0)/ F_A(0)\right]^{1/2}$ from the data. Using $Q^2_{\rm max}=1.0$ GeV$^2$ we find 
\begin{equation}\label{eq:rA}
r_A = 0.81^{+0.05}_{-0.10} \pm {0.14}  \, {\rm fm}  \qquad \mbox{(antineutrino scattering)}, 
\end{equation}
which is consistent with (\ref{eq:MAz}). We conclude that the asymmetry in the errors is not due the inverse relation between $m_A^2$ and $F_A^\prime(0)$. 

The fact that the errors are not symmetric implies that larger values of $m_A$ are acceptable from the fit. Since larger $m_A$ tends to increase the cross section, one can take the asymmetry  to be an indication that the nuclear cross section is too small. To check this option qualitatively, we have multiplied the carbon cross section by a constant factor and repeated the fits. We find that if the factor is larger than 1, the errors indeed become more symmetric. If it is smaller than 1, the asymmetry grows. There is very little change in the value of $m_A^{\rm dipole}$.  We find similar results if we follow the same procedure for the MiniBooNE neutrino data analyzed in \cite{Bhattacharya:2011ah}.  Therefore there are hints that the carbon cross section is too small, which might imply an issue with the nuclear model.  It would be interesting to explore in more detail the combined effect of the $z$ expansion and a change in the nuclear model.  This is left to a future study.

\subsection{Extraction of $\bm{F_A(q^2)}$} 
\begin{figure}
\begin{center}
\includegraphics[scale=1]{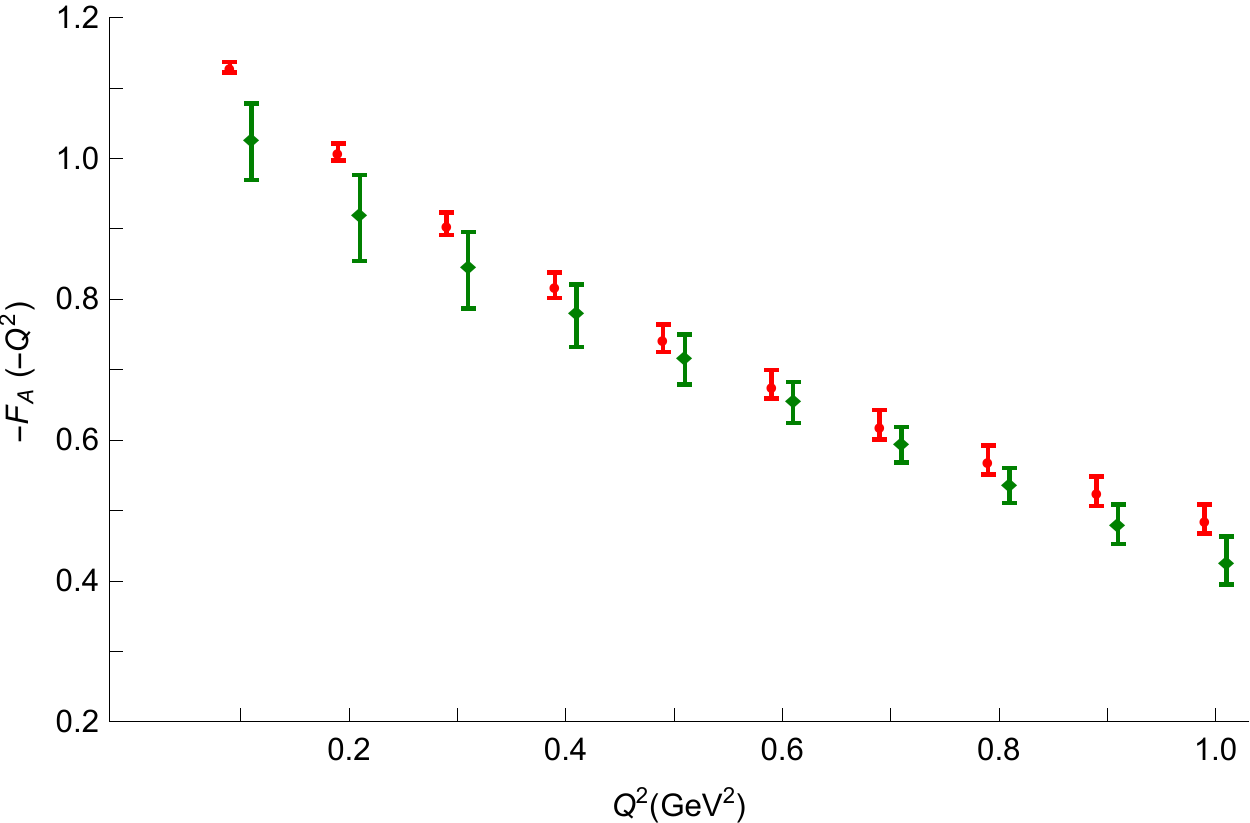}
\caption{\label{fig:FA_Q2} Comparison of the axial-vector form factor $F_A$ as
extracted using the $z$ expansion (green diamonds) 
and dipole ansatz (red circles).
}
\end{center}
\end{figure}

The model-independent approach allows us to extract values of $F_A(q^2)$ from data and not just one parameter such as the axial mass.  We fit $F_A(q^2)$ to the entirety of MiniBooNE's muon antineutrino CCQE data, using the $z$ expansion with $k_{\rm max}=7$ and $|a_k|\le 10$. Figure \ref{fig:FA_Q2} shows the extracted values for $Q^2=0.1,0.2,...,1.0$ GeV$^2$. We compare these results to a dipole fit that assumes $m_A^{\rm dipole}=1.27^{+0.03}_{-0.04}\,{\rm GeV}$. 

\begin{figure}
\begin{center}
\includegraphics[scale=1]{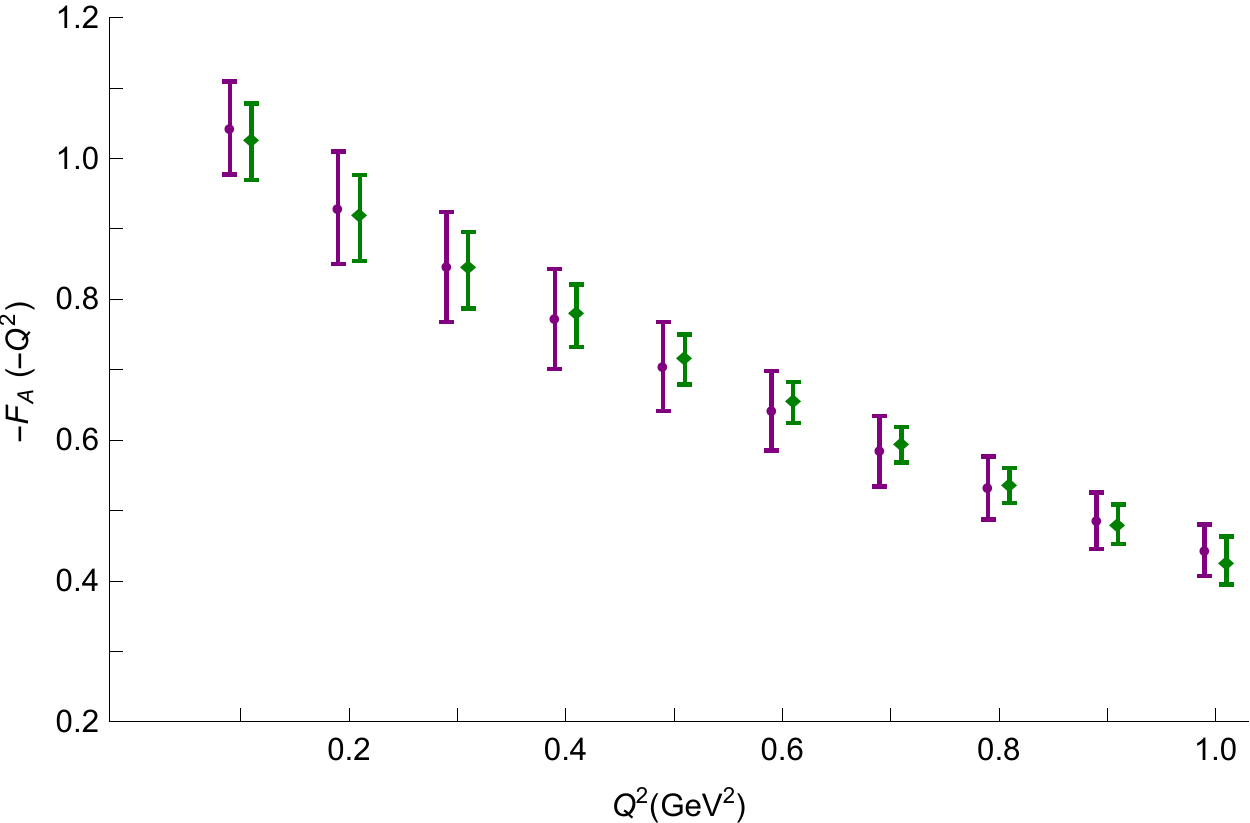}
\caption{\label{fig:nu_vs_nubar} Comparison of extraction of the axial-vector form factor $F_A$ using the $z$ expansion from neutrino data (purple circles) and from antineutrino data (green diamonds). }
\end{center}
\end{figure}

These results can be compared to the extraction of $F_A(q^2)$ in \cite{Bhattacharya:2011ah} from MiniBooNE's muon neutrino CCQE data \cite{AguilarArevalo:2010zc}. Figure \ref{fig:nu_vs_nubar} compares the two extractions.  The very good agreement between the two extractions is clear. 

The shape of the form factor hints at a non-zero curvature. The curvature can be extracted in a model-independent way using the $z$ expansion. In order to do that we extract $a_2$ of equation (\ref{Fz}). We use a cut of  $Q^2_{\rm max}=1.0$ GeV$^2$, $k_{\rm max}=7$ and enforce $|a_k|\le 10$ for $k\ge 3$, i.e. leaving $a_1$ and $a_2$ unbounded (recall that $a_0=F_A(0)=-1.272$). We find  $a_2=-9.6^{+4.7}_{-2.6}$, which can be compared to  $a_2 = -8^{+6}_{-3}$ found in \cite{Bhattacharya:2011ah}. In both cases we see an indication for a non-zero curvature, but it is poorly constrained. A similar extraction of $a_1$, equivalent to the extraction of the axial mass,  gives $a_1=3.4^{+0.9}_{-1.0}$ which is in good agreement with $a_1 = 2.9^{+1.1}_{-1.0}$ of  \cite{Bhattacharya:2011ah}.

One could use the model-independent approach to extract values of $F_A(q^2)$ from other neutrino data sets. These extractions can be used to tabulate values of $F_A(q^2)$ similar to the electromagnetic form factors, see e.g. \cite{Arrington:2007ux}. Such tables would be a much better approach than trying to reconcile various data sets using just one degree of freedom.

\section{Summary}\label{sec:summary}
The improved precision of  experiments require us to move from historical ad-hoc models of form factors to a model-independent approach.  The field of flavor physics have gone through such a change and currently the use of the $z$ expansion is the standard method in analyzing exclusive decays, see e.g. \cite{Aoki:2013ldr}. Motivated by the ``proton radius puzzle" studies of the proton and neutron electromagnetic form factors have started to implement this model-independent method. A similar shift is needed for the use of the axial form factor in neutrino experiments. In particular, one would like to separate effects coming from the nuclear models and effects from the form factors.  The first application of the $z$ expansion to the extraction of the axial mass was performed in \cite{Bhattacharya:2011ah}.  

In this paper we have applied the method of \cite{Bhattacharya:2011ah} to analyze the muon antineutrino CCQE cross section data published by the MiniBooNE experiment \cite{AguilarArevalo:2013hm}. An important difference from the neutrino data analyzed in  \cite{Bhattacharya:2011ah} is that the antineutrino interacts both with protons in carbon and in hydrogen, while the neutrino interacts only with the neutrons in carbon. As a result, in extracting the axial mass, or more generally, the axial form factor, one has to combine cross sections for protons in hydrogen and protons in carbon. While the former is described using form factors alone, the latter requires also the use of a nuclear model. It is important to use the same axial form factor for both. 

We have extracted the axial mass using a model-independent approach from MiniBooNE's muon antineutrino CCQE data \cite{AguilarArevalo:2013hm} . We find $m_A = 0.84^{+0.12}_{-0.04} \pm {0.11}  \, {\rm GeV}$. This result is in very good  agreement with the analogues model-independent extraction of $m_A$ from MiniBooNE's muon neutrino CCQE data \cite{AguilarArevalo:2010zc}, $m_A = 0.85^{+0.22}_{-0.07} \pm 0.09\,{\rm GeV}$ \cite{Bhattacharya:2011ah}. For comparison, a fit using the dipole model gives $m_A^{\rm dipole} = 1.27^{+0.03}_{-0.04}$~GeV, consistent with the value of $m_A^{\rm dipole} = 1.29 \pm 0.05$ reported in  \cite{Bhattacharya:2011ah}.

Our extraction relies on a specific nuclear model, the RFG model, used also by the MiniBooNE experiment. Our study does not address whether this is an adequate model or should it be modified. Since the upper error bar on the model-independent  $m_A$ is about three times larger than the lower error bar, it is possible that some modification of the nuclear model is also needed. It would be interesting to combine the $z$ expansion with other nuclear models.  

 Going beyond a one-parameter comparison, we have extracted values of the axial form factor as a function of $Q^2$. The results are shown in figure \ref{fig:FA_Q2}. These results can be compared to the similar extraction of the axial mass from  MiniBooNE's muon neutrino CCQE data \cite{AguilarArevalo:2010zc}, see figure  \ref{fig:nu_vs_nubar}. We find very good agreement between the two data sets. It would be beneficial to extract values of $F_A$ from other neutrino experiments. Ideally one would like to have a world data set for values of the axial form factor, similar to that of electromagnetic form factors, see e.g. \cite{Arrington:2007ux}
 
In summary, we find very good agreement between MiniBooNE's neutrino and antineutrino data sets when using model-independent extraction of the axial form factor in general and the axial mass in particular.  These model-independent methods should be applied to other data sets and in combination with other nuclear models.  

\vskip 0.2in
\noindent
{\bf Acknowledgements}
\vskip 0.1in
\noindent
We  thank Heather Grebe and Jerold E. Young for collaboration during the early stages of this work. We thank Joe Grange, Richard Hill, Teppei Katori, and Kevin McFarland for useful discussions, and Alexey A. Petrov for his comments on the manuscript. G.P. also thanks the particle physics group of Universit\'e de Montr\'eal, where part of this work was done, for their hospitality. This work was supported by IPP and NSERC of Canada (B.B.),  DOE grant DE-SC0007983 (G.P.), NIST Precision Measurement Grants Program (G.P.), and NSF Grant  PHY-1460853 (A.J.T). 

\begin{appendix}

\section{Appendix: quasielastic (anti)neutrino nucleon scattering cross section} 
The relevant part of the weak-interaction 
Lagrangian is 
\begin{equation}
{\cal L}=\frac{G_F}{\sqrt{2}}V_{ud}\,\bar
{\ell}\gamma^\alpha(1-\gamma_5)\nu\,\bar{u}\gamma_\alpha(1-\gamma_5)d  + {\rm H.c.} \,. 
\end{equation}
The cross section for $\nu(k) + n(p) \to \ell^-(k^\prime)+ p(p^\prime)$ on a free neutron is 
\begin{eqnarray}\label{sigmafree}
\sigma_{\rm free}=\frac1{4|k\cdot p|}\int\frac{d^3{k^\prime}}{(2\pi)^32E_{{\bm k}^\prime}}
\int\frac{d^3{p^\prime}}{(2\pi)^32E_{{\bm p}^\prime}}\overline{\left|{\cal M}^2\right|}(2\pi)^4\delta^4(k+p-k^\prime-p^\prime),
\end{eqnarray}
where the spin-averaged, squared amplitude is 
\begin{eqnarray}
\label{amp2}
\overline{\left|{\cal M}^2\right|}&=&\frac{G_F^2 |V_{ud}|^2 }{4}
L^{\mu\nu} 
\sum_{\rm
  spins}\,\langle p(p^\prime)|\bar{u}\gamma_\mu(1-\gamma_5) d|n(p)\rangle
\langle p(p^\prime)|\bar{u}\gamma_\nu(1-\gamma_5) d|n(p)\rangle^*.
\end{eqnarray}
The leptonic tensor neglecting the neutrino mass is ($\epsilon^{0123}=-1$)
\begin{align}\label{eq:leptonic}
L^{\mu\nu} 
= 
8(k^\mu k^{\prime\nu}+k^\nu k^{\prime\mu}-g^{\mu\nu}k\cdot k^\prime 
- i\epsilon^{\mu\nu\rho\sigma}k_\rho k^{\prime}_{\sigma}) \,.
\end{align} 
The hadronic matrix element appearing in (\ref{amp2}) is parameterized by
\begin{align}
\label{FFdef}
\langle p(p^\prime)|\bar{u} \gamma_\mu(1-\gamma_5) d|n(p)\rangle&=
\bar u^{(p)}(p^\prime) \Gamma_\mu(q) u^{(n)}(p) \,,
\end{align}
where $q=k-k^\prime=p^\prime-p$ and we have defined the vertex function
\begin{multline}
\label{Gamma_mu}
\Gamma_\mu(q) 
= \gamma_\mu F_1(q^2) + {i \over 2m_N}\sigma_{\mu\nu}q^\nu F_2(q^2)+\frac{q_\mu}{m_N}F_S(q^2)
+ \gamma_{\mu}\gamma_5 F_A(q^2) + \frac{p_\mu+p^\prime_\mu}{m_N}\gamma_5F_T(q^2)
\\
+{q_\mu \over m_N} \gamma_5 F_P(q^2)  \,.
\end{multline}
Notice that equations (\ref{FFdef}) and (\ref{Gamma_mu}) define the relative phases between the form factors. In particular, it determines the sign of the ratio of  $F_A(0)$ to $F_1(0)$ measured in neutron decay \cite{Agashe:2014kda}. 

We may write the cross section of (\ref{sigmafree}) as  
\begin{equation}\label{xs}
\sigma_{\rm free} = {G_F^2 |V_{ud}|^2 \over 16 |k\cdot p|} \int{d^3{k^\prime}\over (2\pi)^3 2E_{{\bm k}^\prime}} L^{\mu\nu} \hat{W}_{\mu\nu} \,,
\end{equation}
where the nucleon structure function is 
\begin{equation}
\hat{W}_{\mu\nu} =\int{d^3{p^\prime}\over (2\pi)^32 E_{{\bm p}^\prime}} (2\pi)^4\delta^4(p-p^\prime+q)  H_{\mu\nu} \,.
\end{equation}
The hadronic tensor is
\begin{equation}
H_{\mu\nu} = 
{\rm Tr}[  ( \slash{p}^\prime + m_p )\Gamma_\mu(q) (\slash{p}+m_n )\bar{\Gamma}_\nu(q) ] \,,
\end{equation}
where as usual, $\bar{\Gamma} = \gamma^0 \Gamma^\dagger \gamma^0$.  
We may similarly analyze antineutrino scattering,
$\bar{\nu}(k) + p(p) \to \ell^+(k^\prime) + n(p^\prime)$, 
using (\ref{xs}), taking $L^{\mu\nu} \to L^{\nu\mu}$, 
and making the replacements $m_n \leftrightarrow m_p$, $\Gamma_\mu(q) \to \bar{\Gamma}_\mu(-q)$ in $H_{\mu\nu}$.   

Imposing time-reversal invariance shows that $F_i(q^2)$ are real.    
We will assume isospin symmetry in the following, in which case 
$F_S$ and $F_T$ vanish, $m_n=m_p= m_N$, and $\bar{\Gamma}_\mu(-q)=\Gamma_\mu(q)$.  
The hadronic tensor has the time-reversal invariant decomposition 
\begin{equation}
\label{form1}
H_{\mu\nu} = -g_{\mu\nu} H_1 + {p_\mu p_\nu\over m_N^2} H_2 - i{ \epsilon_{\mu\nu\rho\sigma} \over 2 m_N^2} p^\rho q^\sigma
H_3 + {q_\mu q_\nu \over m_N^2} H_4  
+ {(p_\mu q_\nu + q_\mu p_\nu)\over 2m_N^2 } H_5
\,. 
\end{equation}
The $H_i$'s are expressed in terms of the form factors $F_i$ as 
\begin{eqnarray}
H_1 &=& 8 m_N^2 F_A^2 -2q^2 \left[ (F_1 + F_2)^2 + F_A^2 \right]   
\,,\nonumber \\ 
H_2 &=& H_5 =  8 m_N^2 \left( F_1^2 + F_A^2\right) -2q^2 F_2^2   
\,,\nonumber \\ 
H_3 &=& -16m_N^2\, F_A (F_1 + F_2 ) 
\,,\nonumber \\ 
H_4 &=& -\frac{q^2}{2}\left(F_2^2 + 4F_P^2 \right)-2m_N^2 F_2^2 
- 4m_N^2 \left( F_1F_2 + 2 F_A F_P \right) 
\,.
\end{eqnarray}

In the rest frame of the nucleon, let $E_\ell$ and $|\vec{P}_\ell|=\sqrt{E_\ell^2 - m_\ell^2}$  be the energy and 3-momentum of the charged lepton, and let  $\theta_\ell$ be the angle between the $3$-momenta of the leptons.  Also in that frame $k\cdot p=E_\nu m_N$. Using
\begin{equation}
\int\frac{d^3{p^\prime}}{(2\pi)^32E_{{\bm p}^\prime}}\delta^4(p-p^\prime+q)=\delta(2p\cdot q +q^2), \quad \int\frac{d^3{k^\prime}}{(2\pi)^32E_{{\bm k}^\prime}}=\pi\int \,dE_\ell\,d\cos\theta\,|\vec{P}_\ell|,
\end{equation}
we have 
\begin{multline}\label{xsfree}
\frac{d\sigma_{\rm free}}{dE_\ell d\cos\theta_\ell}=
{G_F^2|V_{ud}|^2 \over 8 \pi  m_N }\delta(2p\cdot q +q^2) |\vec{P}_\ell |
\Bigg\{2 (E_\ell -|\vec{P}_\ell |\cos \theta_\ell)\,H_1
+( E_\ell  + |\vec{P}_\ell |\cos \theta_\ell) H_2\\
\pm \frac{1}{m_N}\Big[(E_\ell -|\vec{P}_\ell |\cos \theta_\ell)(E_\nu+E_\ell )-m^2_\ell \Big]H_3
+ \frac{m_\ell ^2}{m_N^2}(E_\ell -|\vec{P}_\ell |\cos \theta_\ell) H_4- \frac{m_\ell ^2}{m_N}\, H_5\Bigg\}\,,
\end{multline}
where $p\cdot q=m_N(E_\nu-E_\ell)$ and $q^2=m_\ell^2-2E_\nu E_\ell+2E_\nu |\vec{P}_\ell |\cos\theta_\ell$, and 
where the upper (lower) sign is for neutrino (antineutrino) scattering.

We now consider the case of antineutrino-proton scattering. To find the flux-averaged cross section we use the reported antineutrino flux \cite{AguilarArevalo:2013hm} to create a function $f(E_{\bar\nu})$. This function is normalized to one, i.e. $\int dE_{\bar\nu}\,f(E_{\bar\nu})=1$. To obtain the flux averaged cross section,  we write the delta function in (\ref{xsfree}) as 
\begin{equation}\label{E0}
\delta(2p\cdot q+q^2)=\frac{\delta(E_{\bar\nu}-E_0)}{ 2(m_N-E_\ell + |\vec{P}_\ell |\cos\theta)},\, \mbox{ with } 
E_0=\frac{2 E_\ell m_N-m_\ell^2}{2 (m_N-E_\ell + |\vec{P}_\ell |\cos\theta)},
\end{equation}
multiply $d\sigma_{\rm free}$ by $f(E_{\bar\nu})$ and integrate over $E_{\bar\nu}$. 
The flux averaged hydrogen cross section is 
\begin{eqnarray}\label{xsfreeavg}
&&\frac{d\sigma_{\rm hydrogen, avg.}}{dE_\ell d\cos\theta_\ell}=\int dE_{\bar\nu}\,f(E_{\bar\nu})\,\frac{d\sigma_{\rm free}}{dE_\ell d\cos\theta_\ell}=\dfrac{f(E_0)\,G_F^2|V_{ud}|^2 |\vec{P}_\ell |}{ 16 \pi m_N (m_N-E_\ell + |\vec{P}_\ell |\cos\theta)}\times\nonumber\\
&&\Bigg\{2 (E_\ell -|\vec{P}_\ell |\cos \theta_\ell)\,H_1
+( E_\ell  + |\vec{P}_\ell |\cos \theta_\ell) H_2
\pm \frac{1}{m_N}\Big[(E_\ell -|\vec{P}_\ell |\cos \theta_\ell)(E_\nu+E_\ell )-m^2_\ell \Big]H_3
\nonumber\\
&&\quad+ \frac{m_\ell ^2}{m_N^2}(E_\ell -|\vec{P}_\ell |\cos \theta_\ell) H_4- \frac{m_\ell ^2}{m_N}\, H_5\Bigg\}\,.
\end{eqnarray}

The MiniBooNE differential cross section data is given in bins of $\cos\theta$ and $T_\mu=E_\ell-m_l$ with sizes of  0.1 and 0.1 GeV, respectively.  The flux data is given in bins of $E_{\bar\nu}$ of size 0.05 GeV. We use the center of the bin for the cross section data and equation (\ref{E0})  to find $E_0$ and then round it to closest value of $E_{\bar\nu}$ from the center of the bin.  One can also use the exact value of $E_0$ which would change the value of $Q^2$ and $Q^2_{\rm rec}$ in equation (\ref{Q2rec}) but not  $f(E_0)$. We have checked that this different choice has very small effect on the value of the axial mass, well within our reported error bars.   

The carbon scattering cross section per nucleon is given in \cite{Bhattacharya:2011ah} as 
\begin{multline}\label{xscarbon}
\frac{d\sigma_{\rm carbon, per\, nucleon}}{dE_\ell d\cos\theta_\ell}=\frac16
{G_F^2|V_{ud}|^2 |\vec{P}_\ell | \over 16 \pi^2 m_T }
\Bigg\{2 (E_\ell -|\vec{P}_\ell |\cos \theta_\ell)\,W_1
+( E_\ell  + |\vec{P}_\ell |\cos \theta_\ell) W_2
\\
\pm \frac{1}{m_T}\Big[(E_\ell -|\vec{P}_\ell |\cos \theta_\ell)(E_\nu+E_\ell )-m^2_\ell \Big]W_3
+ \frac{m_\ell ^2}{m_T^2}(E_\ell -|\vec{P}_\ell |\cos \theta_\ell) W_4- \frac{m_\ell ^2}{m_T}\, W_5\Bigg\}\,,
\end{multline}
where $W_i$ are given in equation (46) of \cite{Bhattacharya:2011ah}, and the upper (lower) sign is for neutrino (antineutrino) scattering. We have divided by six, the number of neutrons (protons) in carbon, to obtain the cross section per nucleon.  To obtain the flux averaged cross section we multiply by $f(E_{\bar\nu})$ and integrate over $E_{\bar\nu}$ 
\begin{equation}\label{xscarbonavg}
\frac{d\sigma_{\rm carbon, per\, nucleon, avg.}}{dE_\ell d\cos\theta_\ell}=\int dE_{\bar\nu}\,f(E_{\bar\nu})\,\frac{d\sigma_{\rm carbon, per\, nucleon }}{dE_\ell d\cos\theta_\ell}.
\end{equation}
In practice, since $f(E_{\bar\nu})$ is a discrete function, for each bin we use the central value and sum over all of the bins.

\end{appendix}

\end{document}